\documentclass{article}
\RequirePackage{filecontents}
\PassOptionsToPackage{dvipsnames}{xcolor}
\RequirePackage{comgipp}
\RequirePackage{amsfonts}
\RequirePackage{amsmath}
\RequirePackage{authblk}

\usepackage[T1]{fontenc}
\usepackage[utf8]{inputenc}
\usepackage{lmodern}

\usepackage{amsmath}

\usepackage{hyperref}
\usepackage{xcolor}
\hypersetup{
    colorlinks,
    linkcolor={red!50!black},
    citecolor={blue!50!black},
    urlcolor={blue!80!black}
}

\usepackage{graphicx}
\usepackage{adjustbox}

\usepackage{caption}
\usepackage{verbatim}
\usepackage{colortbl}

\usepackage{footmisc}

\newcommand{\theReferenceText}{\fullcite{Scharpf2022b}}
\usepackage[
backend=biber,
style=numeric,
firstinits=true,
url=false,
isbn=false,
safeinputenc,
]{biblatex}
\usepackage{subcaption}

\title{Collaborative and AI-aided Exam Question Generation using Wikidata in Education}
\author[1]{Philipp Scharpf}
\author[2]{Moritz Schubotz}
\author[1]{Andreas Spitz}
\author[3]{André Greiner-Petter}
\author[4]{Bela Gipp}

\affil[1]{University of Konstanz, Germany (\{first.last\}@uni-konstanz.de)}
\affil[2]{FIZ-Karlsruhe, Germany (\{first.last\}@fiz-karlsruhe.de)}
\affil[3]{NII Tokyo, Japan
(greinerpetter@nii.ac.jp)}
\affil[4]{University of Göttingen, Germany (\{last\}@cs.uni-goettingen.de)}
\begin{document}
  \maketitle
  \thispagestyle{firststyle}
  \begin{abstract}
    Since the COVID-19 outbreak, the use of digital learning or education platforms has significantly increased. Teachers now digitally distribute homework and
provide exercise questions. In both cases,
teachers need to continuously develop novel and individual questions. This process can be very time-consuming and should be facilitated and accelerated both through exchange with other teachers and by using Artificial Intelligence (AI) capabilities. To address this need, we propose a multilingual Wikimedia framework that allows for collaborative worldwide teacher knowledge engineering and subsequent AI-aided question generation, test, and correction.
As a proof of concept, we present >>PhysWikiQuiz<<, a physics question generation and test engine. Our system (hosted by Wikimedia at \url{https://physwikiquiz.wmflabs.org}) retrieves physics knowledge from the open community-curated database Wikidata. It can generate questions in different variations and verify answer values and units using a Computer Algebra System (CAS). We evaluate the performance on a public benchmark dataset at each stage of the system workflow. For an average formula with three variables, the system can generate and correct up to 300 questions for individual students based on a single formula concept name as input by the teacher.
  \end{abstract}
\section{Introduction and Motivation}

With the rise of digital learning or education platforms, the frequency of teachers posing tasks and questions digitally has increased substantially.
However, due to temporal constraints, it would be infeasible for teachers to constantly create novel and individual questions tailored to each different student.
With the aid of Artificial Intelligence (AI), they can submit AI-generated learning tests more frequently, which can lead to student performance improvement.
Moreover, many teachers develop exam questions without exchanging ideas or material with their peers. In many cases, this may unnecessarily cost them a lot of time and effort. Instead, they should be able to focus on explaining the concepts to their students.
To address these shortcomings, we propose using Wikidata as a multilingual framework that allows for collaborative worldwide teacher knowledge engineering and subsequent AI-aided question generation, test, and correction.
Using Wikidata in education leads to the research problem need to compare and identify the best-performing methods to generate questions from Wikidata knowledge.

\begin{figure}[t]
    \centering
    \includegraphics[width=0.98\textwidth]{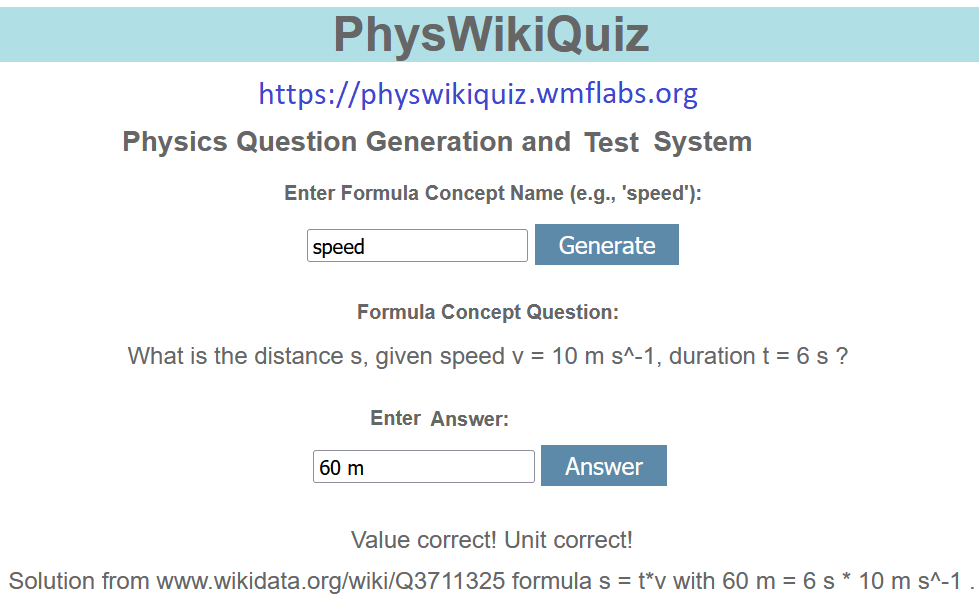}
    \caption{Example question generation (including variable names, symbols, and units) and answer correction (of both solution value and unit) for the formula concept name `speed'. The PhysWikiQuiz system also generates an explanation text with information reference and calculation path.}
    \label{fig:ExampleSpeed}
\end{figure}

As a proof of concept
for the physics domain, we develop and evaluate a >>PhysWikiQuiz<< question generator and solution test engine (example in Figure \ref{fig:ExampleSpeed}), hosted by Wikimedia at \url{https://physwikiquiz.wmflabs.org} with a demovideo available at \url{https://purl.org/physwikiquiz}.
The system addresses the teacher's demand by automatically generating an unlimited number of different questions and values for each student separately. It employs the open access semantic knowledge-base Wikidata\footnote{\url{https://www.wikidata.org}}
to retrieve Wikimedia community-curated physics formulae with identifier (variables with no fixed value\footnote{\url{https://www.w3.org/TR/MathML3/chapter4.html\#contm.ci}}) properties and units using their concept name as input. A given formula is then rearranged, i.e., solved for each occurring identifier by a Computer Algebra System
to create more question sets. For each rearrangement, random identifier values are generated.
Finally, the system compares the student's answer input to a CAS computed solution for both value and unit separately.
%
%
PhysWikiQuiz also provides an API for integration in external education systems or platforms.

To evaluate the system, we pose the following research questions for the assessment of test question generation from Wikidata knowledge (RQs):

\begin{enumerate}
    \item What are the state-of-the-art systems? How to address their shortcomings?
    \item Which information retrieval methods and databases can we employ?
    \item What performance can we achieve?
    \item What are the contributions of the system's modules to this performance?
    \item What challenges occur during implementation and operation?
    \item How can we address these challenges?
\end{enumerate}

\textbf{Structure.} The remainder of this paper is structured as follows. We discuss RQ 1 in Section \ref{sec:Background}, RQs 2-4 in Section \ref{sec:Methods} and \ref{sec:Evaluation}, and RQs 5-6 in Section \ref{sec:Discussion}.

\section{Background and Related Work}\label{sec:Background}


In this section, we review the prerequisite background knowledge
the project builds upon, including the research gap and the employed methods.

\textbf{Question Generation (QG)} is a natural language processing task to generate question-answer (QA) pairs from data (text, knowledge triples, tables, images, and more)\footnote{https://www.microsoft.com/en-us/research/project/question-generation-qg}. The generated QA pairs can then be employed in dialogue systems, such as Question Answering, chatbots, or quizzes.
State of the art is to typically use neural networks to generate structured pairs out of unstructured content extracted from crawled web pages~\cite{DBLP:conf/emnlp/DuanTCZ17}.
There is a number of datasets
and models
openly available with a competitive comparison at \url{https://paperswithcode.com/task/question-generation}.
In the last decade, QA has been increasingly employed and researched for educational applications~\cite{DBLP:conf/icwsm/ChenYHH18,DBLP:conf/acl/SrivastavaG20}. Despite the large variety of techniques, in 2014 only a few had been successfully deployed in real classroom settings~\cite{DBLP:conf/iccsama/LeKP14}.


\textbf{Automated Test Generation (ATG)} for intelligent tutoring systems has so far been tackled using linked open data ontologies to create natural language multiple-choice questions~\cite{DBLP:journals/semweb/VK17}.
%
The evaluation is typically domain-dependent. For example, Jouault et al. conduct a human expert evaluation in the history domain, comparing automatically with manually generated questions to find about 80\% coverage~\cite{jouault2016content}.
%
Some approaches use Wikipedia-based datasets consisting of URLs of Wikipedia articles to generate solution distractors via text similarity~\cite{shah2017automatic}.
%
Since Wikipedia is only semi-structured, it may be more efficient to instead employ highly structured databases. This was attempted by `Clover Quiz', a trivia game powered by DBpedia for multiple-choice questions in English and Spanish.
However, the creators observed the system to have high latency, which is intolerable for a live game. The limitations are addressed by creating questions offline through a data extraction pipeline~\cite{DBLP:journals/semweb/Vega-Gorgojo19}.
%
For the mathematics domain, Wolfram Research released the Wolfram Problem Generator\footnote{\url{https://www.wolframalpha.com/problem-generator}} for AI-generated practice problems and answers. The system covers arithmetics, number theory, algebra, calculus, linear algebra, and statistics, yet is restricted to core mathematics while physics is currently not supported.
Current systems for the physics domain, e.g., `Mr Watts Physics'~\footnote{\url{http://wattsphysics.com/questionGen.html}} and `physQuiz'~\footnote{\url{https://physics.mrkhairi.com}}, are curated only by single maintainers, which leads to a very limited availability of concepts and questions (see Table \ref{tab:ComparisonCompetitors}).



\begin{table}[b]
\resizebox{\columnwidth}{!}{
\begin{tabular}{llll}
\hline
\textbf{System}       & \textbf{Mr Watts Physics} & \textbf{physQuiz Equations} & \textbf{PhysWikiQuiz} \\ \hline
Concepts              & 36                        & 8                           & 469 (Wikidata)        \\ 
Questions per concept & 20                        & 20                          & unlimited             \\ \hline
\end{tabular}}
\caption{Comparison of PhysWikiQuiz scope to competitors.}\label{tab:ComparisonCompetitors}
\end{table}

\textbf{We address the reported shortcomings} by presenting a system for the physics domain that allows for unlimited live question generation from community-curated (Wikidata) knowledge.
Since Wikidata
is constantly growing, our approach scales better than the aforementioned 
static resources curated by single teachers.
Only 2\% of unique concepts were available on `physQuiz Equations' (8 out of 475) and 8\% on `Mr Watts Physics' (36 out of 475), yet 99\% on our `PhysWikiQuiz' (469 out of 475).
In the case of mathematical knowledge, Wikidata currently contains around 5,000 statements that link an item concept name to a formula~\cite{DBLP:conf/semweb/ScharpfSG21}. As stated above, almost 500 of them are from the physics domain.
%
PhysWikiQuiz exploits this information to create, pose, and correct physics questions using mathematical entity linking~\cite{DBLP:conf/www/ScharpfSG21} (in contrast to the competitors), which we review in the following.

\textbf{Mathematical Entity Linking (MathEL)} is the task of linking mathematical formulae or identifiers to unique web resources (URLs), e.g., Wikipedia articles. This requires formula concepts to be identified (first defined and later recognized). For this goal, a `Formula Concept' was defined~\cite{DBLP:conf/sigir/ScharpfSG18,DBLP:conf/sigir/ScharpfSCG19} as a `labeled collection of mathematical formulae that are equivalent but have different representations through notation, e.g., the use of different identifier symbols or commutations'~\cite{DBLP:conf/www/ScharpfSG21}. Formulae appearing in different representations make it difficult for humans and machines to recognize them as instances of the same semantic concept. For example, the formula concept `mass-energy equivalence' can either be written as $E=mc^2$ or $\mu=\epsilon/c^2$ or using a variety of other symbols. To facilitate and accelerate the creation of a large dataset~\cite{DBLP:conf/jcdl/SchubotzGSMCG18} for the training of Formula Concept Retrieval (FCR) methods, a formula and identifier annotation recommender system for Wikipedia articles was developed~\cite{DBLP:conf/recsys/ScharpfMSBBG19}. The FCR approaches are intended to improve the performance of Mathematical Information Retrieval (MathIR) methods, such as Mathematical Question Answering (MathQA)~\cite{mathqajournal,DBLP:conf/clef/ScharpfSGOTG20}, Plagiarism Detection (PD), STEM literature recommendation or classification~\cite{DBLP:conf/jcdl/ScharpfSYHMG20,DBLP:conf/mkm/SchubotzSTKBG20}.

\section{Methods and Implementation}\label{sec:Methods}

In this section, we describe the development of our PhysWikiQuiz physics question generation and test engine, along with the system workflow and module details. PhysWikiQuiz employs the method of Mathematical Entity Linking (see Section \ref{sec:Background}).

The prerequisites for the PhysWikiQuiz system are that it 1) is intended to generate questions as part of an education platform, 2) employs Wikidata as knowledge-base, 3) works on formula concepts, 4) requires formula and identifier unit retrieval, and 5) utilizes a Computer Algebra System to correct the student's answer.

\subsection{System Workflow}



Figure \ref{fig:ExampleSpeed} shows the PhysWikiQuiz User Interface (UI) for an example formula concept name input `speed' with a defining formula of $v=s/t$. The formula can be rearranged as $s=v*t$ or $t=s/v$ (two question sets). For the identifier symbols $v$, $s$, and $t$, their names `velocity', `distance', and `duration' and units `\verb|m s^-1|', `\verb|m|', and `\verb|s|' are retrieved from the corresponding Wikidata item\footnote{\url{https://www.wikidata.org/wiki/Q3711325}}. In the example, the answer is considered as correct in both value `60' and unit `\verb|m|'. If the user clicks again on the `Generate' button, a new question with different formula rearrangement and identifier values is generated. For a system feedback of `Value incorrect!' and/or `Unit incorrect!', the student has the possibility to try other inputs by changing the input field content and clicking again on the `Answer' button.
%
%

The PhysWikiQuiz workflow is divided into six modules (abbreviated by Mx in the following).
In M1, formula and identifier data is retrieved from Wikidata. In M2, the formula is rearranged using the python CAS Sympy\footnote{\label{foot:SymPy}\url{https://www.sympy.org}}. In M3, random values are generated for the formula identifiers. In M4, the question text is generated from the available information. In M5, the student's answer is compared to the system's solution. Finally, M6 generates an explanation text for the student. In case some step or module cannot be successfully executed, the user is notified, e.g., `No Wikidata item with formula found.'
%
%
%
%

\subsection{Modules}
\label{subsec:Modules}


After the user inputs the formula concept name or Wikidata QID (see Figure~\ref{fig:ExampleSpeed}), M1 retrieves the `defining formula' and identifier properties.
%
%
%
PhysWikiQuiz supports all current identifier information formats and strives to stay up to date. The identifier units need to be retrieved from the linked items (in some formats, also the names). Currently, units are stored using the `ISQ dimension' property (P4020 in Wikidata). To make the format more readable for students, the unit strings (e.g., \verb|'L T^-1'|) are translated into SI unit symbols\footnote{\url{https://en.wikipedia.org/wiki/International_System_of_Quantities}} (e.g., \verb|'m s^-1'|).


Having retrieved the required formula and identifier information, M2 is called to generate possible rearrangements using the CAS of SymPy\footref{foot:SymPy}, a python library for symbolic mathematics~\cite{DBLP:journals/cca/JoynerCMG11}.
%
Since the `defining formula' property of the Wikidata item stores the formula in \LaTeX\ format, which is different from the calculable Sympy CAS representation, a translation is necessary. There are several possibilities available for this task.
%
%
The python package LaTeX2Sympy\footnote{\url{https://github.com/OrangeX4/latex2sympy}} is designed to parse \LaTeX\ math expressions and convert it into the equivalent SymPy form.
%
The Java converter LaCASt~\cite{DBLP:journals/aslib/Greiner-PetterS19,greiner2022math}, provided by the VMEXT~\cite{DBLP:conf/mkm/SchubotzMHCG17} API\footnote{\url{https://vmext-demo.formulasearchengine.com/swagger-ui.html}}
translates a semantic \LaTeX\ string to a specified CAS.
In our system evaluation (Section~\ref{sec:Evaluation}), we compare the performance of both translators. For them to work correctly, PhysWikiQuiz performs a number of \LaTeX\ cleanings beforehand, such as replacements and removals that improve the translation performance. 



With the Sympy calculable formula representation available, M3 is ready to replace the right-hand side identifiers with randomly generated integer values. A lower and upper value can be chosen freely. We use the default range from 1 to 10 in our evaluation.
Finally, having successfully replaced the right-hand side identifiers by their respective generated random values, the left-hand side identifier value is calculated. The value is later compared to the student input by M5 (answer correction) to check the validity of the question-answer value.
%
%
At this stage, all information needed to generate a question is available: (1) the formula, (2) the identifier symbols, (3) the identifier (random) values, and (4) the identifier units. M4 generates the question text by inserting the respective information into gaps of a predefined template with placeholders for formula identifier names, symbols, and units.
For a question text example, refer to the screenshot in Figure~\ref{fig:ExampleSpeed}.


After the question text is displayed by the UI, the student can enter an answer consisting of value and unit 
for the left-hand side identifier solution. The information is then parsed by M5. It is subsequently compared to the value output of M1 (solution unit) and M3 (solution value). The student gets feedback on the correctness of value and unit separately.
The system accepts fractions or decimal numbers as input (e.g., $5/2 = 2.5$), which is then compared to the solution with a tolerance that can be specified (default value is $\pm1\%$).
%
%
Finally, after the question is generated and the correctness of the solution is assessed by the system, M6 generates an explanation such that the student can understand how a given solution is obtained. The system returns and displays an explanation text
storing left- and right-hand side identifier names, symbols, values, and units (see M4).
For an explanation text example, refer to the screenshot in Figure \ref{fig:ExampleSpeed}.

\section{Evaluation}\label{sec:Evaluation}

In this section, we present and discuss the results of a detailed PhysWikiQuiz system evaluation at each individual stage of its workflow. We carry out module tests for the individual modules and an integration test to assess the overall performance on a formula concept benchmark dataset (see Section \ref{sec:BenchmarkDataset}). All detailed tables can be found in the \verb|evaluation| folder of the repository\footnote{\label{foot:evalrepo}\url{https://github.com/ag-gipp/PhysWikiQuiz/blob/main/evaluation}}.

\subsection{Benchmark Dataset}
\label{sec:BenchmarkDataset}

The open-access platform `MathMLben'\footnote{\url{https://mathmlben.wmflabs.org/}} stores and displays a benchmark of semantically annotated mathematical formulae~\cite{DBLP:conf/jcdl/SchubotzGSMCG18}. They were extracted from Wikipedia, the arXiv and the Digital Library of Mathematical Functions (DLMF)\footnote{\url{https://dlmf.nist.gov}} and augmented by Wikidata markup~\cite{DBLP:conf/sigir/ScharpfSG18}. The benchmark can be used to evaluate a variety of MathIR tasks, such as the automatic conversion between different CAS~\cite{DBLP:conf/jcdl/SchubotzGSMCG18} or MathQA~\cite{mathqajournal}. The system visualizes the formula expression tree using VMEXT~\cite{DBLP:conf/mkm/SchubotzMHCG17} to reveal how a given formula is processed.
In our PhysWikiQuiz evaluation, we employ a selection of formulae
from the MathMLben benchmark.
The formula concepts were extracted from Wikipedia articles using the formula and identifier annotation recommendation system~\cite{DBLP:conf/recsys/ScharpfMSBBG19,DBLP:conf/www/ScharpfSG21} >>AnnoMathTeX<<\footnote{\label{foot:AMT}\url{https://annomathtex.wmflabs.org}}.

\subsection{Overall System Performance}\label{subsec:SystemPerformance}

Table \ref{tab:IntegrationTest} shows example evaluation results (selection of instances and features) on the MathMLben formula concept benchmark.
For each example concept in the benchmark selection, e.g., `acceleration' (GoldID 310 or Wikidata Q11376), the individual modules are tested individually.
Using a workflow evaluation automation script, we create two separate evaluation tables for the two \LaTeX\ to SymPy translators that we employ (LaTeX2Sympy and LaCASt, see the description of M2 in Section \ref{subsec:Modules}).
%
The overall system performance using the LaTeX2Sympy converter is the following. For 20\% of the concepts, all modules are working properly, and PhysWikiQuiz can provide both a question text, an answer verification with correct internal calculation, and an explanation text. For 29\%, only the question can be displayed, but the system's calculation is wrong, such that the answer correction and explanation text generation do not work correctly. For 52\%, PhysWikiQuiz cannot provide a question.
In summary, the system is able to yield 48\% `question or correction'\footnote{Although the case of `no question but correction' is not very intuitive, it did occur.}, 20\% `question and correction', 29\% question, and 52\% none.
%
The overall system performance using the LaCASt converter is the following. For 26\% of the concepts, all modules are working properly. For 18\%, only the question can be displayed. For 56\%, PhysWikiQuiz can not provide a question.
In summary, the system is able to yield 44\% `question or correction', 26\% `question and correction', 18\% question, 56\% none.

Table \ref{tab:LaTeX2SympyvsLaCASt} summarizes the performance comparison of the two translators.
We include a detailed discussion of the issues in external dependencies that cause this relatively low performance in Section \ref{sec:Discussion}.
Overall, LaCASt performs better in generating both question and correction but cannot provide either question or correction on slightly more instances. We deploy LaCASt in production.

\begin{table}[t]
\centering
\resizebox{\columnwidth}{!}{
\begin{tabular}{lllll}
\hline
\textbf{Translator}  & \textbf{quest. OR corr.} & \textbf{quest. AND corr.} & \textbf{only quest.} & \textbf{none} \\ \hline
\textit{LaTeX2Sympy} & 48\%                     & 20\%                      & 29\%                 & 52\%          \\ 
\textit{LaCASt}      & 44\%                     & 26\%                      & 18\%                 & 56\%          \\ \hline
\end{tabular}}
\caption{Comparison of \textit{LaTeX2Sympy} and \textit{LaCASt} translator in overall system performance for question (quest.) generation and correction (corr.) ability.}
\label{tab:LaTeX2SympyvsLaCASt}
\end{table}

\begin{table}[]
\centering
\resizebox{\columnwidth}{!}{
\begin{tabular}{llllll}
\hline
\textbf{GoldID} & \textbf{QID} & \textbf{Name}        & \textbf{Identifier semantics} & \textbf{Formula translation} & \textbf{Explanation text} \\ \hline
310             & Q11376       & acceleration         & yes                           & no                           & yes                       \\ 
311             & Q186300      & angular acceleration & yes                           & no                           & yes                       \\ 
312             & Q834020      & angular frequency    & yes                           & yes                          & yes                       \\ \hline
\textbf{Total}             &          & \textbf{Performance}         & \textbf{97\% yes}                      & \textbf{60\% yes}                     & \textbf{27\% yes}                  \\
\hline
\end{tabular}}
\caption{Three example evaluation results out of a formula concept selection
from the benchmark MathMLben (\url{https://mathmlben.wmflabs.org/}). Each individual module of the PhysWikiQuiz workflow is evaluated. Here, we only show a summary of the main steps (last three columns condensed from eight, see the
repository).}
\label{tab:IntegrationTest}
\end{table}

\subsection{Module Evaluation}

In the following, we present a detailed evaluation of the individual modules or stages in the workflow.

\paragraph{Retrieval Formula Identifier Semantics and Units}

The first stage of module tests is the assessment of the correct retrieval of the identifier semantics.
Since names and symbols are fetched from Wikidata items that are linked to the main concept item, the retrieval process is prone to errors.
However, we find that for 97\% of the concepts, identifier properties are available in some of the supported formats. 
%

\paragraph{Retrieval of Formula and Identifier Units}

The next workflow stage we evaluate is the formula and identifier unit retrieval.
For 53\% of the test examples, a formula unit is available on the corresponding main concept Wikidata item. For the remaining 47\%, identifier units are available on the respective linked Wikidata items.

\paragraph{LaTeX to SymPy Translation}

The subsequent module tests are concerned with the LaTeX to SymPy translations, which is mandatory for having Sympy rearrange the formula and yield a right-hand side value given random identifier value substitutions (modules 2 and 3).
We evaluate the two converters LaTeX2Sympy and LaCASt in comparison, which were introduced in Section \ref{subsec:Modules}. LaTeX2Sympy is able to yield a correct and calculable SymPy formula in 50\% of the cases. Moreover, it can provide usable Sympy identifiers for the substitutions in 47\% of the cases. For LaCASt, the SymPy formula is correct in 60\% and the SymPy identifiers in 47\%. This means that LaCASt has a better translation performance (10\% more), while the identifier conversion remains the same.

\paragraph{Formula Rearrangement Generation}

Formula rearrangements enhance the availability of additional question variations. In the case of our example `speed', when using Sympy rearrangements, the other variables `distance' and `durations' can also be queried, providing additional concept questions.
%
%
%
%
For lengthy formulae, PhysWikiQuiz can generate a very large amount of question variations. But even for a small formula with 2 identifiers, there are already many possibilities by substituting different numbers as identifier values. On average, the formulae in the test set contain 3 identifiers. Substituting combinations of numbers from 1 to 10, this leads to several hundred potential questions per formula concept.
%
We find that in 27\% of the cases, Sympy can rearrange the `defining formula.' The result is the same for both LaTex2Sympy and LaCASt translation. In comparison to a workflow without M2, 
more than 300 additional questions can be generated.

\paragraph{Right-Hand Side Substitutions and Explanation Text Generation}

The last two module test evaluations assess the success of right-hand side substitutions and explanation text generation.
For LaTeX2Sympy, 45\% of substitutions are made correctly, whereas LaCASt achieves 53\%. Both translators generate correct identifier symbol-value-unit substitutions for the explanation text in 39\% of the test cases.

\section{Discussion}\label{sec:Discussion}

In this section, we discuss our results, contribution, and retrieval challenges of the individual workflow stages and modules. The full list of challenges can be found in the
repository\footref{foot:evalrepo}.

\paragraph{Results and Contribution}

Wikidata currently contains around 5,000 concept items with mathematical formula. Out of these, about 500 are from the physics domain.
Using a Computer Algebra System, PhyWikiQuiz can generate concept questions with value and unit and corrections in around 50\% of the cases.
For a detailed analysis of the errors in the remaining 50\% and a discussion of the challenges to tackle, see the next subsection.

Our contribution is a proof of concept for the physics domain to use Wikidata in education.
We develop a >>PhysWikiQuiz<< question generator and solution test engine and evaluate it on an open formula concept benchmark dataset.
Our work addresses the research gap in comparing methods to generate physics questions from Wikidata knowledge.
We find that using Wikidata and a Computer Algebra System, it is possible to generate an unlimited amount of physics questions for a given formula concept name. Although they all follow a very similar template with very little variation, they contain different variable values, which makes them suitable to provide individual questions for various students.

\subsection{Challenges and Limitations}

\paragraph{Formula Semantics and Translation}


%
%
We manually examine the concepts for which the Wikidata items do not provide units. For some of them, we identify semantic challenges. In our estimation, the concepts either (1) should not have a unit (`ISQ dimension' property) or (2) it is debatable whether they should have one. Example QIDs for the respective cases can be found in the repository\footref{foot:evalrepo}.
In the first case, the respective formulae do not describe physical quantities but formalisms, transformations, systems, or objects. In particular, the formula right-hand side identifier that is calculated does not correspond to the concept item name. In the second case, the corresponding formula provides the calculation of a physical quantity that is not reflected in the concept name. Finally, there is a third case in which the concept item should have a unit property since the formula describes a physical quantity on the right-hand side that is defined by the concept name.
Examining the examples for which the converters cannot provide a properly working translation, we find some challenges that require the development of more advanced \LaTeX\ cleaning methods. Derivative fractions can contain identifier differentiation with or without separating spaces. For example, `acceleration' can be calculated either as \verb|\frac{d v}{d t}| or \verb|\frac{dv}{dt}|. The first formula is correctly translated to the calculable SymPy form \verb|Derivative(v, t)|, whereas the second does not work. Unfortunately, the spaces cannot be introduced automatically in the arguments without losing generality (e.g., \verb|dv| could also mean a multiplication of some identifiers d and v as \verb|d * v|). Implicit multiplication is a general problem. However, it is very likely for a \verb|\frac{}{}| expression with leading \verb|d| symbols in its arguments to contain a derivative, and the risk of losing generality should maybe be taken. In the case of partial derivatives, such as \verb|\frac{\partial v}{\partial t}| the problem does not arise since \verb|\partial| needs a following space to be a proper \LaTeX\ expression.
Some formulae are not appropriate for PhysWikiQuiz question generation and test. The expression \verb|\sum_{i=1}^n m_i(r_i - R) = 0| in `center of mass
' (Q2945123) does not have a single left-hand-side identifier to calculate. The right-hand side is always zero. The equation (correctly) also does not have a formula unit. Finally, expressions like \verb|p_{tot,1} = p_{tot,2}| in `conservation of momentum' (Q2305665) are no functional linkage of identifier variables and thus do not serve as basis for calculation questions.

\paragraph{Identifier Substitutions and Explanation Text Generation}

%
%
For about half of the test examples, the substitution is unsuccessful due to some peculiarities in the defining formula. The full list can be found in the repository\footref{foot:evalrepo}. We encounter the problems that (1) substitutions cannot be made if identifier properties are not available, (2) for some equations, the left-hand side is not a single identifier, but a complex expression or the right-hand side is zero, (3) two equation signs occur in some instances, and (4) identifier properties and formula are not matching in their Wikidata items for some items.
%
%
%
%
The last stage in our workflow evaluation is the assessment of the explanation text correctness. All in all, for 27\% of the concepts, explanation texts can be generated, out of which 39\% contain correct identifier symbol-value-unit substitutions. We conclude that the calculation path display is error-prone and outline some challenges in the following.
%
%
For the explanation texts that are incorrect, we identify some of the potential reasons. We find that (1) in some cases, operators like multiplications are missing, (2) some equations contain dimensionless identifiers, for which the unit is written as the number 1, and (3) in case integrals appear in the formulae, sometimes a mixture of non-evaluated expressions and quantities is displayed.



\subsection{Takeaways}

\paragraph{Answering the research questions.}

%

Having implemented and evaluated the system, we can answer our research question as follows:


\begin{enumerate}
\item PhysWikiQuiz outperforms its competitors by providing a constantly growing number of more than 10 times additional community-curated questions.
    \item We employ and adapt the method of Mathematical Entity Linking
of formula concepts
for question generation using Wikidata.
    \item About 50\% of the benchmark formula Wikidata items can be successfully transformed into questions with correction and explanation. For the remaining cases, we provide an extensive error analysis.
    \item The performance directly depends on formula and identifier name, symbol and unit retrieval, as well as translation to and solving by a CAS.
    \item The bottleneck is caused by the dependencies, such as the CAS Sympy and translator LaCASt. A clearer community agreement on data quality guidelines in Wikidata would also improve the results.
    \item We can improve the quality of the formula cleanliness with user feedback by addressing the issues in the dependencies.
\end{enumerate}

\paragraph{Addressing the challenges.}


To tackle the current limitations, we propose the following solutions:

\begin{itemize}
    \item Formula semantics: Limit use to concepts that can be indisputably associated with formulae and units to avoid unreliability due to community objection.
    \item Formula translation: Increasingly improve the converter performance by receiving and implementing community feedback to enhance concept coverage.
    \item Identifier substitutions: Motivate the Wikidata community to seed the missing identifier properties. This will increase coverage by enabling lacking identifier value substitutions.
    \item Explanation text generation: The problems are expected to be settled with increased data quality of the formula items in Wikidata.
\end{itemize}

Despite the challenges, we have already built an in-production system (with 13 times more coverage than its best-performing competitor) that can and will be used by teachers in practice.

\section{Conclusion and Future Work}


In this paper, we present >>PhysWikiQuiz<<, a physics question generation and test engine. Our system can provide a variety of different questions for a given physics concept, retrieving formula information from Wikidata, correcting the student's answer, and explaining the solution with a calculation path. We separately evaluate each of the six modules of our system to identify and discuss systematic challenges in the individual stages of the workflow. We find that about half of the questions cannot be generated or corrected due to issues that can be addressed by improving the quality of the external dependencies (Wikidata, LaTeX2Sympy, LaCASt, and Sympy) of our system. Our
application demonstrates the potential of mathematical entity linking for education question generation and correction.
%
%

PhysWikiQuiz is listed on the `Wikidata tool pages' for querying data\footnote{\url{https://www.wikidata.org/wiki/Wikidata:Tools/Query_data}}. We welcome the reader to test our system and provide feedback for improvements. If the population of mathematical Wikidata items continues (e.g., by using tools such as >>AnnoMathTeX<<\footref{foot:AMT}), our system will be able to increasingly support additional questions.
We will continue to assess the overall effectiveness of the knowledge transfer from Wikipedia articles to Wikidata items to PhysWikiQuiz questions.
Moreover, we are developing an automation for the Wikidata physics concept item bulk to detect if the question generation or correction is correct, or if the respective items need human edits to make PhysWikiQuiz work.
Detecting these cases will extend the system's operating range and ensures that it works despite the limitations. We also plan to test the system with a larger group of end users.

As a long-term goal, we envision integrating our system into larger education platforms, allowing teachers to simply enter a physics concept about which they want to quiz the students. Students would then receive individually generated questions (via app push notification) on their mobile phones. Having collected all the answers, teachers could then obtain a detailed analysis of the student's strengths and weaknesses and use them to address common mistakes in their lectures. We will evaluate the integrated system with teachers.
%
Finally, we plan to extend our framework with additional question domains, possibly integrating state-of-the-art external dependencies, Wikifunctions\footnote{\url{https://wikifunctions.beta.wmflabs.org}}, and language models as they are developed to increase the coverage further.
With PhyWikiQuiz and its extensions to other educational domains, we hope to make an important contribution to the `Wikidata for Education' project\footnote{\url{https://www.wikidata.org/wiki/Wikidata:Wikidata_for_Education}}.


\section*{Acknowledgment}

This work was supported by the German Research Foundation (DFG grant GI-1259-1).

\printbibliography[keyword=primary]

@inproceedings{DBLP:conf/mkm/SchubotzMHCG17,
  author    = {Moritz Schubotz and
               Norman Meuschke and
               Thomas Hepp and
               Howard S. Cohl and
               Bela Gipp},
  title     = {{VMEXT:} {A} Visualization Tool for Mathematical Expression Trees},
  booktitle = {{CICM}},
  series    = {Lecture Notes in Computer Science},
  volume    = {10383},
  pages     = {340--355},
  publisher = {Springer},
  year      = {2017}
}

@article{DBLP:journals/cca/JoynerCMG11,
  author    = {David Joyner and
               Ondrej Cert{\'{\i}}k and
               Aaron Meurer and
               Brian E. Granger},
  title     = {Open source computer algebra systems: SymPy},
  journal   = {{ACM} Commun. Comput. Algebra},
  volume    = {45},
  number    = {3/4},
  pages     = {225--234},
  year      = {2011}
}

@inproceedings{DBLP:conf/jcdl/SchubotzGSMCG18,
  author    = {Moritz Schubotz and
               Andr{\'{e}} Greiner{-}Petter and
               Philipp Scharpf and
               Norman Meuschke and
               Howard S. Cohl and
               Bela Gipp},
  title     = {Improving the Representation and Conversion of Mathematical Formulae
               by Considering their Textual Context},
  booktitle = {{JCDL}},
  pages     = {233--242},
  publisher = {{ACM}},
  year      = {2018}
}

@inproceedings{DBLP:conf/emnlp/DuanTCZ17,
  author    = {Nan Duan and
               Duyu Tang and
               Peng Chen and
               Ming Zhou},
  editor    = {Martha Palmer and
               Rebecca Hwa and
               Sebastian Riedel},
  title     = {Question Generation for Question Answering},
  booktitle = {Proceedings of the 2017 Conference on Empirical Methods in Natural
               Language Processing, {EMNLP} 2017, Copenhagen, Denmark, September
               9-11, 2017},
  pages     = {866--874},
  publisher = {Association for Computational Linguistics},
  year      = {2017},
  url       = {https://doi.org/10.18653/v1/d17-1090},
  doi       = {10.18653/v1/d17-1090},
  bibsource = {dblp computer science bibliography, https://dblp.org}
}

@inproceedings{DBLP:conf/iccsama/LeKP14,
  author    = {Nguyen{-}Thinh Le and
               Tomoko Kojiri and
               Niels Pinkwart},
  editor    = {Tien Van Do and
               Hoai An Le Thi and
               Ngoc Thanh Nguyen},
  title     = {Automatic Question Generation for Educational Applications - The State
               of Art},
  booktitle = {Advanced Computational Methods for Knowledge Engineering - Proceedings
               of the 2nd International Conference on Computer Science, Applied Mathematics
               and Applications, {ICCSAMA} 2014, 8-9 May, 2014, Budapest, Hungary},
  series    = {Advances in Intelligent Systems and Computing},
  volume    = {282},
  pages     = {325--338},
  publisher = {Springer},
  year      = {2014},
  url       = {https://doi.org/10.1007/978-3-319-06569-4\_24},
  doi       = {10.1007/978-3-319-06569-4\_24},
  bibsource = {dblp computer science bibliography, https://dblp.org}
}

@inproceedings{DBLP:conf/acl/SrivastavaG20,
  author    = {Megha Srivastava and
               Noah Goodman},
  editor    = {Chengqing Zong and
               Fei Xia and
               Wenjie Li and
               Roberto Navigli},
  title     = {Question Generation for Adaptive Education},
  booktitle = {Proceedings of the 59th Annual Meeting of the Association for Computational
               Linguistics and the 11th International Joint Conference on Natural
               Language Processing, {ACL/IJCNLP} 2021, (Volume 2: Short Papers),
               Virtual Event, August 1-6, 2021},
  pages     = {692--701},
  publisher = {Association for Computational Linguistics},
  year      = {2021},
  url       = {https://doi.org/10.18653/v1/2021.acl-short.88},
  doi       = {10.18653/v1/2021.acl-short.88},
  bibsource = {dblp computer science bibliography, https://dblp.org}
}

@inproceedings{DBLP:conf/icwsm/ChenYHH18,
  author    = {Guanliang Chen and
               Jie Yang and
               Claudia Hauff and
               Geert{-}Jan Houben},
  title     = {LearningQ: {A} Large-Scale Dataset for Educational Question Generation},
  booktitle = {Proceedings of the Twelfth International Conference on Web and Social
               Media, {ICWSM} 2018, Stanford, California, USA, June 25-28, 2018},
  pages     = {481--490},
  publisher = {{AAAI} Press},
  year      = {2018},
  url       = {https://aaai.org/ocs/index.php/ICWSM/ICWSM18/paper/view/17857},
  bibsource = {dblp computer science bibliography, https://dblp.org}
}

@inproceedings{DBLP:conf/www/ScharpfSG21,
  author    = {Philipp Scharpf and
               Moritz Schubotz and
               Bela Gipp},
  title     = {Fast Linking of Mathematical Wikidata Entities in Wikipedia Articles
               Using Annotation Recommendation},
  booktitle = {{WWW} (Companion Volume)},
  pages     = {602--609},
  publisher = {{ACM} / {IW3C2}},
  year      = {2021}
}

@inproceedings{DBLP:conf/recsys/ScharpfMSBBG19,
  author    = {Philipp Scharpf and
               Ian Mackerracher and
               Moritz Schubotz and
               J{\"{o}}ran Beel and
               Corinna Breitinger and
               Bela Gipp},
  title     = {\emph{AnnoMath TeX} - a formula identifier annotation recommender
               system for {STEM} documents},
  booktitle = {RecSys},
  pages     = {532--533},
  publisher = {{ACM}},
  year      = {2019}
}

@inproceedings{DBLP:conf/jcdl/ScharpfSYHMG20,
  author    = {Philipp Scharpf and
               Moritz Schubotz and
               Abdou Youssef and
               Felix Hamborg and
               Norman Meuschke and
               Bela Gipp},
  title     = {Classification and Clustering of arXiv Documents, Sections, and Abstracts,
               Comparing Encodings of Natural and Mathematical Language},
  booktitle = {{JCDL}},
  pages     = {137--146},
  publisher = {{ACM}},
  year      = {2020}
}

@inproceedings{DBLP:conf/mkm/SchubotzSTKBG20,
  author    = {Moritz Schubotz and
               Philipp Scharpf and
               Olaf Teschke and
               Andreas K{\"{u}}hnemund and
               Corinna Breitinger and
               Bela Gipp},
  title     = {AutoMSC: Automatic Assignment of Mathematics Subject Classification
               Labels},
  booktitle = {{CICM}},
  series    = {Lecture Notes in Computer Science},
  volume    = {12236},
  pages     = {237--250},
  publisher = {Springer},
  year      = {2020}
}

@article{mathqajournal,
  author    = {Moritz Schubotz and
               Philipp Scharpf and
               Kaushal Dudhat and
               Yash Nagar and
               Felix Hamborg and
               Bela Gipp},
  title     = {Introducing MathQA - {A} Math-Aware Question Answering System},
  journal   = {Information Discovery and Delivery},
  volume    = {42, No. 4},
  pages     = {214--224},
  year      = {2019},
  doi       = {10.1108/IDD-06-2018-0022},
}

@inproceedings{DBLP:conf/sigir/ScharpfSG18,
  author    = {Philipp Scharpf and
               Moritz Schubotz and
               Bela Gipp},
  title     = {Representing Mathematical Formulae in Content MathML using Wikidata},
  booktitle = {BIRNDL@SIGIR},
  series    = {{CEUR} Workshop Proceedings},
  volume    = {2132},
  pages     = {46--59},
  publisher = {CEUR-WS.org},
  year      = {2018}
}

@inproceedings{DBLP:conf/sigir/ScharpfSCG19,
  author    = {Philipp Scharpf and
               Moritz Schubotz and
               Howard S. Cohl and
               Bela Gipp},
  title     = {Towards Formula Concept Discovery and Recognition},
  booktitle = {BIRNDL@SIGIR},
  series    = {{CEUR} Workshop Proceedings},
  volume    = {2414},
  pages     = {108--115},
  publisher = {CEUR-WS.org},
  year      = {2019}
}

@article{DBLP:journals/aslib/Greiner-PetterS19,
  author    = {Andr{\'{e}} Greiner{-}Petter and
               Moritz Schubotz and
               Howard S. Cohl and
               Bela Gipp},
  title     = {Semantic preserving bijective mappings for expressions involving special
               functions between computer algebra systems and document preparation
               systems},
  journal   = {Aslib J. Inf. Manag.},
  volume    = {71},
  number    = {3},
  pages     = {415--439},
  year      = {2019}
}

@article{DBLP:journals/semweb/VK17,
  author    = {Vinu E. V and
               P. Sreenivasa Kumar},
  title     = {Automated generation of assessment tests from domain ontologies},
  journal   = {Semantic Web},
  volume    = {8},
  number    = {6},
  pages     = {1023--1047},
  year      = {2017},
  url       = {https://doi.org/10.3233/SW-170252},
  doi       = {10.3233/SW-170252},
  bibsource = {dblp computer science bibliography, https://dblp.org}
}

@article{jouault2016content,
  title={Content-dependent question generation using LOD for history learning in open learning space},
  author={Jouault, Corentin and Seta, Kazuhisa and Hayashi, Yuki},
  journal={Transactions of the Japanese Society for Artificial Intelligence},
  pages={LOD--F},
  year={2016},
  publisher={The Japanese Society for Artificial Intelligence}
}

@inproceedings{shah2017automatic,
  title={Automatic question generation for intelligent tutoring systems},
  author={Shah, Riken and Shah, Deesha and Kurup, Lakshmi},
  booktitle={2017 2nd International Conference on Communication Systems, Computing and IT Applications (CSCITA)},
  pages={127--132},
  year={2017},
  organization={IEEE}
}

@article{DBLP:journals/semweb/Vega-Gorgojo19,
  author    = {Guillermo Vega{-}Gorgojo},
  title     = {Clover Quiz: {A} trivia game powered by DBpedia},
  journal   = {Semantic Web},
  volume    = {10},
  number    = {4},
  pages     = {779--793},
  year      = {2019},
  url       = {https://doi.org/10.3233/SW-180326},
  doi       = {10.3233/SW-180326},
  bibsource = {dblp computer science bibliography, https://dblp.org}
}

@inproceedings{DBLP:conf/clef/ScharpfSGOTG20,
  author    = {Philipp Scharpf and
               Moritz Schubotz and
               Andr{\'{e}} Greiner{-}Petter and
               Malte Ostendorff and
               Olaf Teschke and
               Bela Gipp},
  title     = {ARQMath Lab: An Incubator for Semantic Formula Search in zbMATH Open?},
  booktitle = {{CLEF} (Working Notes)},
  series    = {{CEUR} Workshop Proceedings},
  volume    = {2696},
  publisher = {CEUR-WS.org},
  year      = {2020}
}

@inproceedings{DBLP:conf/semweb/ScharpfSG21,
  author    = {Philipp Scharpf and
               Moritz Schubotz and
               Bela Gipp},
  title     = {Mathematics in Wikidata},
  booktitle = {Wikidata@ISWC},
  series    = {{CEUR} Workshop Proceedings},
  volume    = {2982},
  publisher = {CEUR-WS.org},
  year      = {2021}
}

@article{greiner2022math,
  title={Do the Math: Making Mathematics in Wikipedia Computable},
  author={Greiner-Petter, Andr{\'e} and Schubotz, Moritz and Breitinger, Corinna and Scharpf, Philipp and Aizawa, Akiko and Gipp, Bela},
  journal={IEEE Transactions on Pattern Analysis and Machine Intelligence},
  year={2022},
  publisher={IEEE}
}

@InProceedings{Scharpf2022b,
  title = {Collaborative and AI-aided Exam Question Generation using Wikidata in Education},
  booktitle = {Proceedings of the 3rd Wikidata Workshop (Wikidata 2022) co-located with the 21th International Semantic Web Conference (ISWC 2022)},
  author = {Scharpf, Philipp and Schubotz, Moritz and Spitz, Andreas and Greiner-Petter, Andre and Gipp, Bela},
  publisher = {{CEUR} Workshop Proceedings},
  year = {2022},
  month = {October},
  location = {Hangzhou, China (virtual)},
  doi = {10.13140/RG.2.2.30988.18568},
  topic = {mathir},
}
\end{document}